\documentclass[11pt]{article}

\usepackage{fullpage}

\usepackage{cite}
\usepackage{amsmath,amssymb,amsfonts}
\usepackage{algorithm}
\usepackage[noend]{algpseudocode}
\usepackage{graphicx}
\usepackage{textcomp}
\usepackage{xcolor}
\usepackage{setspace}
\usepackage{tabularx,multirow,colortbl,booktabs}
\usepackage{graphicx,subcaption}
\usepackage{geometry}
\usepackage{authblk}

\newcommand{\abs}[1]{\lvert #1 \rvert}

\def\BibTeX{{\rm B\kern-.05em{\sc i\kern-.025em b}\kern-.08em
    T\kern-.1667em\lower.7ex\hbox{E}\kern-.125emX}}
\begin{document}

\renewcommand\Authands{ and }

\newcommand*{\email}[1]{%
  \normalsize\href{mailto:#1}{#1}\par
}

\title{\vspace{-1ex}\LARGE\textbf{Routing-Led Placement of VNFs in Arbitrary Networks}\footnote{This manuscript is currently under peer review for possible publication. The reviewer can use this version interchangeably.}}

\author[1]{\normalsize Joseph Billingsley}
\author[1]{\normalsize Ke Li}
\author[1]{\normalsize Wang Miao}
\author[1]{\normalsize Geyong Min}
\author[2]{\normalsize Nektarios Georgalas}
\affil[1]{\normalsize Department of Computer Science, University of Exeter, EX4 4QF, Exeter, UK}
\affil[2]{\normalsize Research and Innovation, British Telecom, Adastral Park, Martlesham, UK}
\affil[$\ast$]{\normalsize Email: \texttt{jb931@exeter.ac.uk}}

\date{}

\maketitle

\vspace{-3ex}
{\normalsize\textbf{Abstract: }	The ever increasing demand for computing resources has led to the creation of hyperscale datacentres with tens of thousands of servers. As demand continues to rise, new technologies must be incorporated to ensure high quality services can be provided without the damaging environmental impact of high energy consumption. Virtualisation technology such as network function virtualisation (NFV) allows for the creation of services by connecting component parts known as virtual network functions (VNFs). VNFs cam be used to maximally utilise available datacentre resources by optimising the placement and routes of VNFs, to maintain a high quality of service whilst minimising energy costs. Current research on this problem has focussed on placing VNFs and considered routing as a secondary concern. In this work we argue that the opposite approach, a routing-led approach is preferable. We propose a novel routing-led algorithm and analyse each of the component parts over a range of different topologies on problems with up to 16000 variables and compare its performance against a traditional placement based algorithm. Empirical results show that our routing-led algorithm can produce significantly better, faster solutions to large problem instances on a range of datacentre topologies.}


\section{Introduction}

The internet and the services it has enabled have changed how we work \cite{OECD16}, travel \cite{ZhengMF15}, socialise \cite{Bargh04} and even how we engage with politics \cite{Farrell12}, government \cite{ChadwickM13} and the rest of society \cite{Castells14}. The infrastructure that provides these services is invisible to most people. One critical part of this infrastructure is the datacentre which provide the computing power needed to provide services. Demand for datacentre resources has increased year-on-year and this trend is expected to continue \cite{AvgerinouBC17}. New technology such as augmented and virtual reality \cite{OrloskyKT17}, the internet of things \cite{PalattellaDGRTE16} and smart cities \cite{ChourabiNWGMNPS12} may have further positive impacts on society and businesses but will demand more from the underlying infrastructure. Existing techniques cannot meet these new demands without significant and damaging environmental impacts \cite{AndraeE15}. Instead, we must develop solutions that can efficiently use existing resources to allow for the continued development of high quality services whilst minimising the energy consumption of the datacentres that provide them.

Resolving these two conflicting objectives requires a comprehensive solution that integrates recent hardware and software innovations. Innovations in the structure of the datacentre aim to improve its scalability. These techniques include new network topologies, the arrangement of switches and servers that allow servers to communicate with each other, that facilitate the addition of equipment to the datacentre. Innovation is also occuring at the software level. Previously, common tasks such as firewalls and deep packet inspection would be executed by specialised equipment known as network functions. In a virtualised datacentre these tasks are provided by Virtual Network Functions (VNFs), software running on virtual machines. VNFs can be created, moved and destroyed far faster than physical hardware.

In a modern virtualised datacentre, a service is provided by placing and connecting sequences of VNFs to form a service. The quality of a service (QoS) is determined
by metrics such as latency - the expected time for a packet to traverse the service - and packet loss - the proportion of packets that do not complete the service due to an overuse of certain servers or switches. The QoS of each service must be maximised whilst also minimising the datacentre energy consumption. A solution to this problem defines the placement of VNFs and routes between VNFs to form a set of services that meet these conflicting objectives simultaneously. This is known as the VNF placement problem (VNFPP) and has been shown to be NP-Hard~\cite{LuizelliCBG17}.

Due to the NP-Hardness of the problem, and the very high number of servers in modern datacentres, several heuristic and metaheuristic solutions have been proposed \cite{LuizelliCBG17, CaoZACHS16,FonsecaF98, LeivadeasFLIK18, RankothgeMLRL15,JiangLHCC12,BariCAB15,LangeGZTJ17,GouarebFA18,BillingsleyLMMG19}. Typically, these approaches place a set of VNFs and then find routes between VNFs. These algorithms implicitly optimise routes by finding improving placements. In this work we categorise these approaches as \textit{placement-led} optimisation techniques. However, key metrics such as latency, packet loss and energy are often minimised by using the shortest, least congested routes available \cite{BillingsleyLMMG19}. Hence we should allow the routes to dictate the placement. We categorise these as \textit{routing-led} optimisation of the VNFPP.

To be able to solve large scale problems, routing-led algorithm must be able to efficiently find nearby servers that can accommodate each VNF in the service. Some existing works can be classified as routing led. Some use knowledge of the structure of the graph to simplify placement \cite{BillingsleyLMMG19, ChuaWZSH16} which prevents the solution from solving arbitrary graphs. Others have considered all possible servers and selected the most suitable~\cite{BariCAB15, LangeGZTJ17} but this strategy is unsuitable for large datacentres. The work by Goureb \textit{et al}~\cite{GouarebFA18} considered three possible routes between the first and last VNFs of a service and placed VNFs only on these routes. This does not guarantee a VNF will be placed when there is a server with sufficient capacity.

This paper proposes a novel metaheuristic to tackle the VNFPP at scale on arbitrary graphs. Our major contributions are as follows:

\begin{itemize}
    \item We evaluate three strategies to select suitable servers in a scalable routing-led placement algorithm
    \item We propose a routing-led genetic algorithm approach to the VNFPP that uses these strategies to optimise large scale problems
    \item We compare our algorithm against a representative placement-led algorithm and evaluate their effectiveness on multiple network topologies with up to 16,000 servers.
\end{itemize}

The remainder of this paper is organised as follows. Section \ref{sec:literature_review} reviews the current work on the VNFPP and on comparable metaheuristics in particular. Section \ref{sec:problem_formulation} provides a formal definition of the VNFPP and the multiobjective formulation used in this work. Section \ref{sec:optimisation} describes the genetic algorithm and operators considered in this work including alternative server location strategies. In Section \ref{sec:evaluation} the effectiveness of each component of the algorithm is tested and our solution is evaluated against a comparable placement-led approach. Finally, Section \ref{sec:conclusion} concludes this paper and outlines some potential future directions.

\section{Literature Review}
\label{sec:literature_review}
The VNF placement problem requires solving two problems simultaneously: VNFs must be placed and routes must be found to form services.

In many works linear programming techniques have been used \cite{MiottoLCG19, AddisBBS15, JemaaPP16, GaoABS18, OljiraGTB17, BaumgartnerRB15, AllegKMA17, MarottaZDK17} to solve both of these problems simultaneously. However current implementations can only consider problem instances with hundreds of servers. For larger problem instances heuristic and metaheuristic approaches are more suitable. The work by Luizelli \textit{et al}~\cite{LuizelliCBG17} proposed to combine variable neighbourhood search with a linear programming solver. VNS selects a subset of variables that are considered by the linear programming solver whilst the other variables remain fixed. This approach solves the largest problem instances we are aware of in the literature but it is still time consuming requiring 10,000 seconds (\texttildelow2.5 hours) to solve a problem instance with 1000 servers. Cao \textit{et al}~\cite{CaoZACHS16} used NSGA-II~\cite{DebAPM02} and MOGA~\cite{FonsecaF98} with a binary matrix solution representation which represented both placement and routing instructions. In this solution representation most of the search space is infeasible and hence custom initialisation and repair operators were used to repair invalid solutions.

Some works do not consider the routing directly and instead consider it as a function of the placement. Leivadeas \textit{et al}~\cite{LeivadeasFLIK18} use a Tabu search metaheuristic to place each VNF. The shortest routes between VNFs in a service are then used. Similarly, Rankothge \textit{et al}~\cite{RankothgeMLRL15} considered problems with up to 400 VNFs, using a Genetic Algorithm with problem specific operators to place VNFs and a depth first search heuristic to find routes between VNFs in a service. In \cite{JiangLHCC12},, Jiang \textit{et al} used a local search heuristic to find improving solutions that only require one VNF migration. A greedy search heuristic then considers new positions for the VNF and favours positions that can be accessed with low congestion routes. In each of these works, the placement determines the routing. The algorithm implicitly optimises routes by finding improving placements.

An alternative approach is to allow the routing to decide the placement of VNFs. One way to apply this principle is to represent the possible placements of VNFs as nodes in a multistage graph where each edge has some cost assigned to it determined by the route between the two current and next VNF. A solution to the problem is a path through the graph. Bari \textit{et al}~\cite{BariCAB15} first proposed this representation and used a dynamic programming heuristic to solve VNFPP problem instances, aiming to minimise the `traffic forwarding cost' calculated based on the distance between two nodes and the quantity of traffic that was sent between them. Lange \textit{et al}~\cite{LangeGZTJ17} utilised the same representation and costs but utilised the Pareto Simulated Annealing algorithm and modified the neighbourhood generating step. In their variant, neighbouring solutions are generated by probabilistically exploring the multistage graph with a greater chance of selecting paths that minimise cost. Notably, both of these works consider every possible server at each step. On large problems with many servers this can be a time consuming operation.

%
%
%

More recently, Gouareb \textit{et al}~\cite{GouarebFA18} utilised a similar routing led approach in their work. The authors assumed the starting and end points of the service are fixed and found the three shortest routes between these two points. They then optimise the placement of the VNFs for each service whilst ensuring they can only be placed on nodes on these paths. Our earlier work on the VNF Placement Problem used a similar approach~\cite{BillingsleyLMMG19}. We utilised the layout of the fat tree topology to have a greater likelihood of placing VNFs on the nearest servers. However this technique is topology specific and will not apply to arbitrary graphs.

Most existing work on the VNF placement problem is limited to solving small scale problem instances. Scalable heuristic and metaheuristic approaches that consider the placement directly have been proposed which implicitly optimise routes between VNFs. Existing routing-led solutions to the VNF placement problem require fixed starting or end points and use techniques that have not been shown to scale to large problems. This work provides a scalable routing-led solution to the VNF placement problem without the limitations of earlier works.

\section{Problem Formulation}
\label{sec:problem_formulation}
In the NFV placement problem, the network is represented as a graph, $G = (N, E)$ where $N$ denotes the set of network nodes and $E$ represents the set of links connecting the nodes. The node in the graph can be either a physical network device, e.g. switches, router, gateway, or a server capable of running virtual machines. Let $S_i$ be the $i$th server in the topology. Multiple VNFs form a service chain to provide a service. Let $V_{i,j}$ denote the $j$th VNF in the $i$th service chain. Multi-path routing strategies, e.g. equal-cost multi-path routing (ECMP), are used in datacentre networks to improve transmission efficiency and reliability. Let a path be a sequence of graph nodes connected by edges. A solution to the VNF placement problem is an assignment of VNFs to server nodes and a collection of paths for each service. Let $P_{V_i}^k$ be the $kth$ path of the $ith$ service in the solution. Let $A_{S_i}$ denote the VNFs assigned to the $i$th server. Finally, let $C_{S_i}$ be the available capacity of the $i$th server and $C_{V_{i,j}}$ be the required capacity of the $j$th VNF of the $i$th service.

The placement and routing parts of the solution are connected by their constraints: (1) the required capacity of the VNFs assigned to a server cannot exceed the available capacity and (2) an instance of each VNF in the service must be visited in the route. Formally these constraints can be defined as:

\begin{enumerate}
    \item $\sum_{\forall v \in A_i} C_v \leq C_{S_i}$
    \item $\forall v \in V_i, \exists s \in P_{V_i}^k$ s.t. $ A_s \supseteq v$
\end{enumerate}

Several models exist to evaluate the quality of a solution to the VNFPP. In this work we will use a queueing theory model proposed in \cite{BillingsleyLMMG19} and extended in \cite{BillingsleyJOURNAL} that estimates the expected latency and packet loss of each service and the total energy consumption of the datacentre. As these objectives can conflict, we formulate this as a multi-objective optimisation problem with three objectives: the average service latency, the average service packet loss and the total energy consumption of the datacentre.

\section{Routing-Led Optimisation of the VNFPP}
\label{sec:optimisation}
A routing-led optimisation of the VNFPP aims to place sequential VNFs of a service on nodes with short paths between them. In this work, we design a set of operators to be used with multiobjective genetic algorithm algorithms. We first design a solution representation and operators that optimise the placement of the first VNFs of each service. Next we propose three routing-led server selection strategies that use information about the network topology to place the remaining VNFs in each service. Finally we find a set of routes between the VNFs for a service based on the ECMP routing protocol to form a complete solution. 

\subsection{Service Origin Operators}
\label{sec:service_optimisation}
The building block hypothesis of genetic algorithms \cite{ForrestM92} indicates that good solutions are constructed from good `building blocks' and hence dependent components should be exchanged during crossover. In the VNFPP the best service quality is achieved by selecting paths that are not being utilised by other services. However due to limited datacentre resources and energy concerns, it is not typically possible for each service to have dedicated resources. It is more likely that two services that start from closer servers will use the same nodes and hence share resources. Good solutions to the VNFPP will minimise resource conflicts between services whilst balancing service quality and energy consumption. We should therefore design our solution representation and operators so that the distance between services is maintained during crossover.

In hierarchical networks such as the Fat Tree or Leaf-Spine datacentre topologies (see Fig. \ref{fig:topologies}) servers can be encoded as a string whilst maintaining the relative distances between servers. However, this only applies to a subset of topologies. For example there is no string encoding of the Dcell representation that maintains the relative distances between all servers. Variable length string (VLS) representations and operators also use strings to encode solutions but are designed to find and exchange related groups of variables during optimisation \cite{GoldbergKD89}. Rather than specifying a suitable solution representation for each topology a VLS representation may find suitable groups of services during the optimisation.

As we are considering a range of topologies, we experiment with both of these solution representations. In the fixed length string representation, each server is represented by an array that contains the ID of the services that should originate from there. In the VLS representation, each character is a tuple of the service ID and the server ID it should originate from. For a fair comparison we use similar genetic operators for both representations: single point crossover and uniform mutation for the fixed length string representation and messy crossover \cite{GoldbergKD89} for the VLS. Uniform initialisation is used in both cases. Server IDs are assigned as shown in Fig. \ref{fig:topologies}. Notably this arrangement of IDs ensures some distance information is maintained in the fixed length string representation.

\subsection{Server Selection Strategies}
\label{sec:opt_expansion}
To evaluate a solution, the remaining VNFs in the service must be placed. Each VNF should be placed on the nearest available server with sufficient capacity. This operation must be efficient as it will be performed many times per solution evaluation. We evaluate three different approaches to this problem a simple breadth first search algorithm, a cached variant which stores information on distance and a novel spanning tree based strategy.

\subsubsection{Simple BFS}
A breadth first search (BFS) will progressively consider the closest nodes. The search procedure can stop as soon as a server is found with enough capacity and has a worst case time complexity of $O(\abs{N} + \abs{E})$.

\subsubsection{Cached BFS}
A BFS will expand the same servers many times during optimisation. Instead a BFS can be performed from every server before the optimisation begins and the sequence of nearest servers can be stored for each server. This has a preprocessing time complexity of $O((\abs{N} + \abs{E}) \cdot \abs{S})$ operations and a memory complexity of $O(\abs{S}^2)$. Subsequent operations have a worst case time complexity of $O(\abs{S})$.

\subsubsection{Spanning Tree Search}
On very large networks executing a BFS for every service is wasteful and the preprocessing time and memory requirements of the `Cached' algorithm may be very high. We propose an alternative technique that combines the simple search and cached approaches to find feasible solutions quickly with lower preprocessing time and memory complexity.

\begin{algorithm}[t]

    \setstretch{1.1}

    \caption{Update spanning tree search tables}

    \begin{algorithmic}[1]
        \Procedure{Update}{\textit{ST}, \textit{start}, \textit{capacity}}
        \State let $Q$ be a queue
        \State label \textit{start} as discovered
        \State $Q$.enqueue((\textit{start}$, $\textit{start}$, $\textit{capacity}$,0$))
        \While{$Q$ is not empty}
        \State $(v, p, c, d) \gets Q$.dequeue()
        \State $v$.rowFor($p$).cap $\gets c$ \Comment Set row capacity
        \State $v$.rowFor($p$).dist $\gets d$ \Comment Set row distance
        \State find the best ($b$) and second best ($s$) rows in $v$
        \If{the best row has changed}
        \For{$n$ in $ST$.adjacentEdges($v$)}
        \If{$n$ is discovered}
        \State \textbf{continue}
        \EndIf

        \State label $n$ as discovered

        \If{$n = b$.edge}
        \State $Q$.enqueue($n,v,b$.cap$,b$.dist$+1$)
        \Else
        \State $Q$.enqueue($n,v,s$.cap$,s$.dist$+1$)
        \EndIf
        \EndFor
        \ElsIf{the second best row has changed}
        \If{$b$.edge is not discovered}
        \State label $b$.edge as discovered
        \State $Q$.enqueue($n,v,s$.cap$,s$.dist$+1$)
        \EndIf
        \EndIf
        \EndWhile
        \EndProcedure
    \end{algorithmic}
    \label{alg:update_spanning}
\end{algorithm}

In this technique the network topology graph is first converted into a spanning tree $ST$. A table is constructed for each node in the spanning tree which tracks the distance to and the capacity of the best server for each edge. where the `best' server is the nearest server with sufficient capacity. The tables are initialised by performing a BFS from each server over the spanning tree. When a node is added to the search front the algorithm checks whether the source server is better than the server currently recorded for the current edge and updates the edge if this is true. If a table is not updated the search does not need to propagate to this nodes children. This has a worst case time complexity of $O((\abs{N} + \abs{E}) \cdot \abs{S})$.

\begin{table*}[!t]
    \caption{Time taken preprocessing, preparing a solution for evaluation and total for 10,000 solutions}
    \label{tbl:expansion_results}
    \begin{tabularx}{\textwidth}{Xlrrrrrrrr}
        \toprule
                   & \multirow{2}{*}{$\abs{S}$} & \multicolumn{2}{c}{Preprocessing (s)} & \multicolumn{3}{c}{Mean Solution Preparation (ms)} & \multicolumn{3}{c}{Total for 10000 Solutions (s)}                                                                                                                                             \\
                   &                            & Cached                                & Spanning                                         & Simple                                            & Cached                   & Spanning                  & Simple                     & Cached                   & Spanning                   \\
        \cmidrule{1-2} \cmidrule(l){3-4} \cmidrule(l){5-7} \cmidrule(l){8-10}
                   & 420                        & \cellcolor{gray!25} 0.00              & \cellcolor{gray!25} 0.00                         & 0.07                                              & \cellcolor{gray!25} 0.02 & 0.14                      & 0.69                       & \cellcolor{gray!25} 0.20 & 1.38                       \\
                   & 930                        & \cellcolor{gray!25} 0.00              & \cellcolor{gray!25} 0.00                         & 0.17                                              & \cellcolor{gray!25} 0.05 & 0.29                      & 1.71                       & \cellcolor{gray!25} 0.52 & 2.94                       \\
        Dcell      & 8190                       & 373.00                                & \cellcolor{gray!25} 0.00                         & 1.76                                              & \cellcolor{gray!25} 0.62 & 2.09                      & \cellcolor{gray!25} 17.55  & 379.22                   & 20.90                      \\
                   & 17556                      & 3586.00                               & \cellcolor{gray!25} 0.00                         & 3.99                                              & \cellcolor{gray!25} 1.19 & 4.69                      & \cellcolor{gray!25} 39.88  & 3597.86                  & 46.94                      \\
                   & 33306                      & ---                                   & \cellcolor{gray!25} 1.00                         & \cellcolor{gray!25} 8.06                          & ---                      & 9.30                      & \cellcolor{gray!25} 80.61  & ---                      & 93.97                      \\
                   & 74256                      & ---                                   & \cellcolor{gray!25} 5.00                         & \cellcolor{gray!25} 21.68                         & ---                      & 22.96                     & \cellcolor{gray!25} 216.83 & ---                      & 234.60                     \\
        \cmidrule{1-2} \cmidrule(l){3-4} \cmidrule(l){5-7} \cmidrule(l){8-10}
                   & 432                        & \cellcolor{gray!25} 0.00              & \cellcolor{gray!25} 0.00                         & 0.27                                              & \cellcolor{gray!25} 0.02 & 0.10                      & 2.70                       & \cellcolor{gray!25} 0.21 & 1.00                       \\
                   & 1024                       & \cellcolor{gray!25} 0.00              & \cellcolor{gray!25} 0.00                         & 1.14                                              & \cellcolor{gray!25} 0.05 & 0.23                      & 11.37                      & \cellcolor{gray!25} 0.53 & 2.34                       \\
        Fat Tree   & 8192                       & 209.00                                & \cellcolor{gray!25} 0.00                         & 35.75                                             & \cellcolor{gray!25} 0.76 & 2.16                      & 357.48                     & 216.56                   & \cellcolor{gray!25} 21.64  \\
                   & 16000                      & 1361.00                               & \cellcolor{gray!25} 0.00                         & 83.99                                             & \cellcolor{gray!25} 1.17 & 3.55                      & 839.86                     & 1372.65                  & \cellcolor{gray!25} 35.54  \\
                   & 27648                      & ---                                   & \cellcolor{gray!25} 0.00                         & 168.30                                            & ---                      & \cellcolor{gray!25} 6.93  & 1682.96                    & ---                      & \cellcolor{gray!25} 69.31  \\
                   & 65536                      & ---                                   & \cellcolor{gray!25} 2.00                         & 403.05                                            & ---                      & \cellcolor{gray!25} 18.88 & 4030.51                    & ---                      & \cellcolor{gray!25} 190.80 \\
        \cmidrule{1-2} \cmidrule(l){3-4} \cmidrule(l){5-7} \cmidrule(l){8-10}
                   & 392                        & \cellcolor{gray!25} 0.00              & \cellcolor{gray!25} 0.00                         & 0.16                                              & \cellcolor{gray!25} 0.02 & 0.09                      & 1.58                       & \cellcolor{gray!25} 0.19 & 0.86                       \\
                   & 1152                       & \cellcolor{gray!25} 0.00              & \cellcolor{gray!25} 0.00                         & 0.67                                              & \cellcolor{gray!25} 0.06 & 0.26                      & 6.66                       & \cellcolor{gray!25} 0.63 & 2.59                       \\
        Leaf-Spine & 8192                       & 111.00                                & \cellcolor{gray!25} 0.00                         & 12.26                                             & \cellcolor{gray!25} 0.60 & 1.93                      & 122.59                     & 117.00                   & \cellcolor{gray!25} 19.30  \\
                   & 15488                      & 717.00                                & \cellcolor{gray!25} 0.00                         & 36.93                                             & \cellcolor{gray!25} 1.16 & 3.86                      & 369.25                     & 728.62                   & \cellcolor{gray!25} 38.61  \\
                   & 32768                      & ---                                   & \cellcolor{gray!25} 0.00                         & 143.32                                            & ---                      & \cellcolor{gray!25} 9.74  & 1433.18                    & ---                      & \cellcolor{gray!25} 97.40  \\
                   & 64082                      & ---                                   & \cellcolor{gray!25} 1.00                         & 503.74                                            & ---                      & \cellcolor{gray!25} 22.39 & 5037.42                    & ---                      & \cellcolor{gray!25} 224.87 \\
        \bottomrule
    \end{tabularx}

    \vspace{1ex}
\end{table*}

To place a VNF the placement algorithm greedily takes the nearest edge with sufficient capacity. As a spanning tree is fully connected and has no loops, exactly one edge for each node will lead to a server with the maximum capacity and each edge will lead to a distinct server. Additionally, as the algorithm always takes steps towards the closest server, the placement step will never revisit a node. Hence it is clear that the algorithm will always find a server that can accommodate the current VNF if such a server exists.

Once a VNF has been placed we must update all tables that reference the changed server to correct their capacity. To do so we propagate information on the current best server the node is aware of from the changed node to adjacent nodes recursively. Similar to the construction process we can perform this efficiently with a BFS as a node must only propagate information if its best or second best row was changed as a result of the update. As the `best' server for a table has the highest capacity and is the closest, tables can have different best nodes which minimises the number of updates needed. In particular, most messages will not propagate past the graphs \textit{central nodes}, the nodes which minimise the greatest distance to all other vertices \cite{SmartS99}. Due to its distance to other nodes, central node will hold a reference to the best servers in the datacentre. Hence an update will only propagate past this node if it was from one of the best servers in the datacentre. In the worst case, placement will traverse the diameter of the spanning tree. If an update does pass a central node all tables will need to be updated as all tables will hold a reference to the changed server. Hence the worst case time complexity for the expansion is $O(d + \abs{N} + \abs{E})$ where $d$ is the diameter of the network. Pseudocode for the update step is given in Alg. \ref{alg:update_spanning}.

\subsection{Routing}
To simplify the generation of solutions, the expansion step outputs a sequence of VNF placements but does not dictate how packets should be routed between them. As in ECMP, in this work traffic is split over the set of shortest routes between nodes. This information can be stored efficiently by creating a set of condensed routing tables for each node in the datacentre topology. Each node stores the range of server IDs that are on the shortest path of each edge. A server ID can appear in multiple ranges if multiple shortest routes exist. The set of paths for the solution can be found by recursively following the instructions in the routing tables.


    





\section{Performance Evaluation}
\label{sec:evaluation}

\begin{table*}[!t]
    \caption{Mean hypervolume for each algorithm after 10000 evaluations}
    \label{tbl:optimisation_results}
    \setlength\tabcolsep{5pt}
    \begin{tabularx}{\textwidth}{Xlp{0.1cm}rp{0.1cm}rp{0.1cm}rp{0.1cm}rp{0.1cm}rp{0.1cm}rp{0.1cm}rp{0.1cm}rp{0.1cm}r}
        \toprule
                                    & \multirow{2}{*}{$\abs{S}$} & \multicolumn{6}{c}{Placement Led} & \multicolumn{6}{c}{Fixed Length String} & \multicolumn{6}{c}{Variable Length String}                                                                                                                                                                                                                                                                                                                     \\
                                    &                            & \multicolumn{2}{c}{IBEA}          & \multicolumn{2}{c}{NSGA-II}             & \multicolumn{2}{c}{MOEA/D}                 & \multicolumn{2}{c}{IBEA} & \multicolumn{2}{c}{NSGA-II} & \multicolumn{2}{c}{MOEA/D} & \multicolumn{2}{c}{IBEA} & \multicolumn{2}{c}{NSGA-II} & \multicolumn{2}{c}{MOEA/D}                                                                                                                                         \\
        \cmidrule{1-2} \cmidrule(l){3-8} \cmidrule(l){9-14} \cmidrule(l){15-20}
                                    & 420                        & \textdagger                       & 0.298                                   & \textdagger                                & 0.298                    & \textdagger                 & 0.300                      &                          & 0.755                       &                            & \cellcolor{gray!25} 0.766 &             & 0.742 & \textdagger & 0.706                     & \textdagger & 0.699 & \textdagger & 0.691 \\
        Dcell                       & 930                        & \textdagger                       & 0.302                                   & \textdagger                                & 0.303                    & \textdagger                 & 0.301                      &                          & 0.745                       &                            & \cellcolor{gray!25} 0.751 & \textdagger & 0.725 & \textdagger & 0.715                     & \textdagger & 0.709 & \textdagger & 0.708 \\
                                    & 8190                       & \textdagger                       & 0.337                                   & \textdagger                                & 0.337                    & \textdagger                 & 0.337                      &                          & 0.733                       &                            & 0.739                     & \textdagger & 0.728 &             & \cellcolor{gray!25} 0.742 & \textdagger & 0.728 &             & 0.737 \\
        \cmidrule{1-2} \cmidrule(l){3-8} \cmidrule(l){9-14} \cmidrule(l){15-20}
                                    & 432                        & \textdagger                       & 0.419                                   & \textdagger                                & 0.417                    & \textdagger                 & 0.403                      &                          & 0.661                       &                            & \cellcolor{gray!25} 0.667 & \textdagger & 0.639 & \textdagger & 0.621                     & \textdagger & 0.602 & \textdagger & 0.606 \\
        \multirow{2}{*}{Fat Tree}   & 1024                       & \textdagger                       & 0.457                                   & \textdagger                                & 0.457                    & \textdagger                 & 0.453                      &                          & 0.695                       &                            & \cellcolor{gray!25} 0.701 & \textdagger & 0.670 & \textdagger & 0.670                     & \textdagger & 0.651 & \textdagger & 0.658 \\
                                    & 8192                       & \textdagger                       & 0.511                                   & \textdagger                                & 0.511                    & \textdagger                 & 0.511                      &                          & 0.728                       &                            & \cellcolor{gray!25} 0.734 & \textdagger & 0.719 &             & 0.734                     & \textdagger & 0.711 &             & 0.726 \\
                                    & 16000                      & \textdagger                       & ---                                     & \textdagger                                & ---                      & \textdagger                 & ---                        &                          & 0.749                       &                            & \cellcolor{gray!25} 0.749 & \textdagger & 0.740 & \textdagger & 0.741                     & \textdagger & 0.717 & \textdagger & 0.733 \\
        \cmidrule{1-2} \cmidrule(l){3-8} \cmidrule(l){9-14} \cmidrule(l){15-20}
                                    & 392                        & \textdagger                       & 0.597                                   & \textdagger                                & 0.593                    & \textdagger                 & 0.573                      &                          & 0.823                       &                            & \cellcolor{gray!25} 0.832 &             & 0.813 & \textdagger & 0.784                     & \textdagger & 0.756 & \textdagger & 0.764 \\
        \multirow{2}{*}{Leaf-Spine} & 1152                       & \textdagger                       & 0.573                                   & \textdagger                                & 0.574                    & \textdagger                 & 0.571                      &                          & 0.804                       &                            & \cellcolor{gray!25} 0.809 & \textdagger & 0.778 & \textdagger & 0.777                     & \textdagger & 0.752 & \textdagger & 0.768 \\
                                    & 8192                       & \textdagger                       & 0.562                                   & \textdagger                                & 0.562                    & \textdagger                 & 0.562                      &                          & 0.779                       &                            & \cellcolor{gray!25} 0.784 & \textdagger & 0.768 &             & 0.781                     & \textdagger & 0.749 & \textdagger & 0.767 \\
                                    & 15488                      & \textdagger                       & ---                                     & \textdagger                                & ---                      & \textdagger                 & ---                        &                          & 0.791                       &                            & \cellcolor{gray!25} 0.792 & \textdagger & 0.778 & \textdagger & 0.785                     & \textdagger & 0.740 & \textdagger & 0.772 \\
        \bottomrule
    \end{tabularx}

    \vspace{1ex}
    {\raggedright The highest mean hypervolumes before rounding are highlighted. \textdagger\ Denotes highlighted result is significantly better according to a Wilcoxon's ranked sum test,, $P < 0.05$.}
\end{table*}

Each algorithm was evaluated on three popular datacentre topologies: Fat Tree~\cite{Al-FaresLV08}, Leaf-Spine~\cite{Cisco19} and Dcell~\cite{GuoWTSZL08}. Each of these datacentre topologies were designed for a different use case. The Fat Tree topology is designed to allow for equal bandwidth on all routes in large datacentres using commodity switches. The Leaf-Spine topology is intended for datacentres with high inter-server traffic. The Dcell topology is designed to be robust against failures. Fat Tree and Leaf-Spine are hierarchical datacentre topologies whilst Dcell is a server centric topology. Groups of 12 tests were run in parallel using a CPU with 12 threads and 6 cores with a 3.4GHz base clock and 32Gb of RAM.

\subsection{Server Selection Strategies}
To evaluate the efficiency of each server selection strategy 1000 different solutions were generated for each network topology at six different scales from $\sim$400 servers to $\sim$70,000 servers. Services were added to each problem instance until at least 60\% of available capacity was required to place all services. The average preprocessing and processing times are listed in Table \ref{tbl:expansion_results}. The largest instances of the test problem for the `cached' selection strategy required more memory than was available for testing and have been ommitted. 

As the cached strategy does not directly use the network topology after preprocessing, solution preparation time is comparable over each datacentre topology. Additionally the cached technique places services 4-5x faster than the other strategies. However on large scale problems the preprocessing step is resource intensive, in particular memory intensive, which makes it unsuitable for large problem instances.

In contrast, the simple strategy is significantly affected by the choice of network topology. The simple selection strategy is able to solve large problem instances on the Dcell network topology in an acceptable amount of time but is an order of magnitude slower on both hierarchical topologies. As the simple selection strategy will stop as soon as it finds a server with sufficient capacity it benefits from densely connected server centric topologies where most steps will lead to a server. On hierarchical topologies such as Leaf-Spine and Fat Tree a BFS originating from a server will visit all switches in the network before it considers most of the servers.

The spanning strategy produces consistent results across each topology and can efficiently scale to large problem instances. However it underperforms when compared to the cached strategy on smaller problems. As the spanning strategy must perform a BFS to update node tables after placing each VNF, its performance is comparable to the simple server selection strategy on problems with densely connected servers such as the Dcell. However on hierarchical problems the spanning strategy can stop earlier more often as it is likely to stop at a central node.

As the preprocessed datacentre topology can be used for many tests, the cached selection strategy may be a suitable choice for small problem instances, problems where many function evaluations will be used or situations with very large amounts of resources. The simple strategy may be a good choice for placement algorithms designed to solve server centric topologies. For general purpose problems the spanning tree algorithm produces consistently good results and can scale to large problem instances.

\begin{table*}[!t]

    \caption{Mean execution time for each algorithm after 10000 evaluations}
    \label{tbl:optimisation_times}
    \setlength\tabcolsep{5pt}
    \begin{tabularx}{\textwidth}{Xlp{0.1cm}rp{0.1cm}rp{0.1cm}rp{0.1cm}rp{0.1cm}rp{0.1cm}rp{0.1cm}rp{0.1cm}rp{0.1cm}r}
        \toprule
                                    & \multirow{2}{*}{$\abs{S}$} & \multicolumn{6}{c}{Placement Led} & \multicolumn{6}{c}{Fixed Length String} & \multicolumn{6}{c}{Variable Length String}                                                                                                                                                                                                                                                                                                               \\
                                    &                            & \multicolumn{2}{c}{IBEA}          & \multicolumn{2}{c}{NSGA-II}             & \multicolumn{2}{c}{MOEA/D}                 & \multicolumn{2}{c}{IBEA} & \multicolumn{2}{c}{NSGA-II} & \multicolumn{2}{c}{MOEA/D} & \multicolumn{2}{c}{IBEA} & \multicolumn{2}{c}{NSGA-II} & \multicolumn{2}{c}{MOEA/D}                                                                                                                                   \\
        \cmidrule{1-2} \cmidrule(l){3-8} \cmidrule(l){9-14} \cmidrule(l){15-20}
                                    & 420                        & \textdagger                       & 14                                      & \textdagger                                & 15                       & \textdagger                 & 20                         & \textdagger              & 8                           &                            & 7                       & \textdagger & 10   & \textdagger & 8    &             & \cellcolor{gray!25} 6    & \textdagger & 10   \\
        Dcell                       & 930                        & \textdagger                       & 63                                      & \textdagger                                & 67                       & \textdagger                 & 98                         &                          & \cellcolor{gray!25} * 22    &                            & \cellcolor{gray!25} 22  & \textdagger & 32   & \textdagger & 24   &             & \cellcolor{gray!25} 22   & \textdagger & 38   \\
                                    & 8190                       & \textdagger                       & 9577                                    & \textdagger                                & 11073                    & \textdagger                 & 16437                      &                          & \cellcolor{gray!25} 2826    & \textdagger                & 4093                    & \textdagger & 4295 & \textdagger & 3726 &             & 3054                     & \textdagger & 6544 \\
        \cmidrule{1-2} \cmidrule(l){3-8} \cmidrule(l){9-14} \cmidrule(l){15-20}
                                    & 432                        & \textdagger                       & 36                                      & \textdagger                                & 43                       & \textdagger                 & 47                         & \textdagger              & 9                           &                            & \cellcolor{gray!25} * 7 & \textdagger & 16   & \textdagger & 9    &             & \cellcolor{gray!25} 7    & \textdagger & 13   \\
        \multirow{2}{*}{Fat Tree}   & 1024                       & \textdagger                       & 143                                     & \textdagger                                & 172                      & \textdagger                 & 144                        & \textdagger              & 23                          &                            & 21                      & \textdagger & 54   & \textdagger & 25   &             & \cellcolor{gray!25} 20   & \textdagger & 50   \\
                                    & 8192                       & \textdagger                       & 6461                                    & \textdagger                                & 6209                     & \textdagger                 & 5542                       &                          & 555                         & \textdagger                & 717                     & \textdagger & 1526 & \textdagger & 630  &             & \cellcolor{gray!25} 460  & \textdagger & 1734 \\
                                    & 16000                      &                                   & ---                                     &                                            & ---                      &                             & ---                        & \textdagger              & 2983                        & \textdagger                & 4229                    & \textdagger & 4799 & \textdagger & 2696 & \textdagger & \cellcolor{gray!25} 2650 & \textdagger & 7009 \\
        \cmidrule{1-2} \cmidrule(l){3-8} \cmidrule(l){9-14} \cmidrule(l){15-20}
                                    & 392                        & \textdagger                       & 19                                      & \textdagger                                & 23                       & \textdagger                 & 25                         & \textdagger              & 7                           &                            & \cellcolor{gray!25} * 5 & \textdagger & 11   & \textdagger & 7    &             & \cellcolor{gray!25} 5    & \textdagger & 10   \\
        \multirow{2}{*}{Leaf-Spine} & 1152                       & \textdagger                       & 115                                     & \textdagger                                & 123                      & \textdagger                 & 102                        &                          & 22                          &                            & \cellcolor{gray!25} 21  & \textdagger & 48   & \textdagger & 25   &             & 23                       & \textdagger & 48   \\
                                    & 8192                       & \textdagger                       & 3972                                    & \textdagger                                & 3925                     & \textdagger                 & 3547                       &                          & 476                         & \textdagger                & 744                     & \textdagger & 1174 & \textdagger & 634  &             & \cellcolor{gray!25} 451  & \textdagger & 1671 \\
                                    & 15488                      &                                   & ---                                     &                                            & ---                      &                             & ---                        & \textdagger              & 2161                        & \textdagger                & 3082                    & \textdagger & 3630 & \textdagger & 2209 & \textdagger & \cellcolor{gray!25} 1591 & \textdagger & 6607 \\
        \bottomrule
    \end{tabularx}

    \vspace{1ex}
    {\raggedright The highest mean times are highlighted. \textdagger\ Denotes highlighted result is significantly better according to a Wilcoxon's ranked sum test, $P < 0.05$. \par * Indicates which setting was used for significance tests }
\end{table*}

\subsection{Routing-Led Optimisation}
The routing-led algorithms were next evaluated against a representative placement-led one. To provide a fair comparison, genetic operators were selected that are similar to those used in the fixed length string (FLS) routing-led algorithm. In the initialisation step, the initial solutions are placed as in the FLS initialisation operator but the remaining VNFs are assigned to subsequent servers in the string. The same mutation and crossover operators are used for the placement-led algorithm as in the routing-led algorithms. Without a repair mechanism the placement-led algorithm, mutation and crossover rarely generate feasible solutions. A repair mechanism is applied that moves VNFs on overfilled servers to the nearest server in the string that can accommodate it. Paths are constructed as in the FLS operators but VNFs are selected by their distance in the string.

Each algorithm was used to solve 30 problem instances for each datacentre topology at different scales. The operators were used with the multiobjective evolutionary algorithms: NSGA-II \cite{DebAPM02}, IBEA \cite{ZhangL07}, and MOEA/D \cite{ZitzlerK04}. For each problem instance a set of services was generated that require at least 60\% of the available capacity. In both the FLS and VLS representation, the cached server selection strategy was used to prepare a solution. The hypervolume (HV) metric was used to compare the overall results. As larger networks will host more services and hence use more energy we consider the expected energy contribution of each service rather than the total energy in the HV calculation. The nadir and utopian points were estimated using the worst and best values for each objective for all solutions and all objectives were normalised before calculating the HV. Table \ref{tbl:optimisation_results} shows the mean hypervolume from 30 problem instances after 10,000 iterations and Table \ref{tbl:optimisation_times} shows the execution time, excluding preprocessing. Memory constraints prevented the problem for being run on larger problem instances. Results on the largest instances of the Dcell and for the larger instances for the placement-led algorithm are not unavailable as these tests did not terminate in an acceptable amount of time.

In all instances the routing-led algorithms significantly outperformed the placement-led algorithm. These results support the argument that routes should lead the placement of VNFs. Placing VNFs in a routing-led strategy increases the chance that the VNFs in each service are near to each other. A placement-led strategy does not have this benefit and requires the evolutionary algorithm to improve routes via improving placements.

Notably the FLS representation significantly outperformed the VLS representation. As a string representation can still maintain some information on the distance between servers, it is likely that the FLS may result in good initial building blocks for the crossover. In comparison, the uniform initialisation used for the VLS does not reflect the relationship between servers immediately.

Results from Table \ref{tbl:optimisation_times} show the routing-led algorithm strategy was significantly faster than the placement-led strategy in this instance. As the placement-led algorithm has not been optimised for this problem it is not a fair comparison. However, the routing-led algorithms are three orders of magnitude faster than other metaheuristics that have considered problems of this scale in the literature (22-38 seconds vs 10,000 seconds for problems with \texttildelow1000 servers \cite{LuizelliCBG17}) and the problems considered are significantly larger than have been considered in the literature so far. As the field has not yet converged on a problem definition for the VNFPP it is not possible to compare the solution qualities directly.

Notably, Dcell requires an order of magnitude longer to optimise than other topologies on larger topologies despite having a similar number of servers and switches to other topologies. It is possible that this is due to the arrangement of the topology information in memory. The Fat Tree and Leaf-Spine topologies can be laid out in memory such that nearby servers in the topology are also near to each other in memory. This is not true for a Dcell topology for the same reasons a string representation cannot fully capture information on the distance between servers discused in Section \ref{sec:optimisation}. As a result memory accesses in the Dcell topology are less predictable than memory accesses in hierarchical topologies. This factor should be considered in the design of future models and metaheuristics.

\section{Conclusions and Future Work}
\label{sec:conclusion}
Existing metaheuristic works on the VNF placement problem typically aim to place VNFs without considering the datacentre topology. An alternative approach places VNFs in a service so that they are joined by short routes. We classify this group of algorithms as routing-led techniques. Existing works that meet this classification require fixed starting and ending locations or only worked for specific topologies. In this work we proposed a new routing-led optimisation algorithm that does not have these limitations. The algorithm was tested against multiple topologies with datacentres containing up to 16,000 servers and 1000s of services. The routing-led algorithm was found to be significantly faster and produce better results than a comparable placement based algorithm on all test instances and shown to scale to problems much larger than have previously been considered.

The algorithm is able to solve large scale problem instances efficiently on most of the topologies considered. Future work could consider cache-friendly or cache oblivious evaluation techniques to allow the optimisation process to efficiently solve large scale graphs with unpredictable memory accesses. Many other recent EMO algorithms can be considered in future~\cite{LiKCLZS12,ChenLBY18,ZouJYZZL19,LiZZL09,LiZLZL09,Li19,LiCSY19,WuLKZZ19,LiK14,LiFK11,LiKWTM13,CaoKWL12,CaoKWL14,LiKZD15,LiDZZ17,LiKD15,ChenLY18,LiZKLW14,LiFKZ14,LiKWCR12,LiWKC13,CaoKWLLK15,LiCFY19,WuLKZ20,WuKJLZ17,WuLKZZ17,LiDY18,WuKZLWL15,LiDZK15}. Additionally alternative solution representations that may better capture information on service quality should be considered.

\section*{Acknowledgment}
K. Li was supported by UKRI Future Leaders Fellowship (Grant No. MR/S017062/1) and Royal Society (Grant No. IEC/NSFC/170243). J. Billingsley was supported by EPSRC Industrial CASE and British Telecom (Grant No. 16000177).

\bibliographystyle{IEEEtran}
\bibliography{IEEEabrv,bibliography}

\end{document}